\documentclass[reprint,english,aip,apl,bibnotes]{revtex4-1}
\usepackage[latin9]{inputenc}
\usepackage{amstext}
\usepackage{graphicx}
\usepackage{amssymb}
\usepackage{multirow}
\usepackage{dcolumn}
\usepackage{color}
\usepackage{amsmath}
\usepackage{babel}
\usepackage{epstopdf}
\usepackage{natbib}
\usepackage{hyperref}
\usepackage{sidecap}
\definecolor{BLUE}{rgb}{0,0,1}
\hypersetup{
     colorlinks   = true,
     citecolor    = BLUE,
     urlcolor = BLUE,
     linkcolor = black
}

\makeatletter

\makeatletter
\begin{document}
\title{A Method for Building Low Loss Multi-Layer Wiring for Superconducting Microwave Devices}

\author{A. Dunsworth}
\affiliation{Department of Physics, University of California, Santa Barbara, California 93106-9530, USA}

\author{R. Barends}
\affiliation{Google Inc., Santa Barbara, CA 93117, USA}

\author{Yu Chen}
\affiliation{Google Inc., Santa Barbara, CA 93117, USA}

\author{Zijun Chen}
\affiliation{Department of Physics, University of California, Santa Barbara, California 93106-9530, USA}

\author{B. Chiaro}
\affiliation{Department of Physics, University of California, Santa Barbara, California 93106-9530, USA}

\author{A. Fowler}
\affiliation{Google Inc., Santa Barbara, CA 93117, USA}

\author{B. Foxen}
\affiliation{Department of Physics, University of California, Santa Barbara, California 93106-9530, USA}

\author{E. Jeffrey}
\affiliation{Google Inc., Santa Barbara, CA 93117, USA}

\author{J. Kelly}
\affiliation{Google Inc., Santa Barbara, CA 93117, USA}

\author{P. V. Klimov}
\affiliation{Google Inc., Santa Barbara, CA 93117, USA}

\author{E. Lucero}
\affiliation{Google Inc., Santa Barbara, CA 93117, USA}

\author{J.Y. Mutus}
\affiliation{Google Inc., Santa Barbara, CA 93117, USA}

\author{M. Neeley}
\affiliation{Google Inc., Santa Barbara, CA 93117, USA}

\author{C. Neill}
\affiliation{Department of Physics, University of California, Santa Barbara, California 93106-9530, USA}

\author{C. Quintana}
\affiliation{Google Inc., Santa Barbara, CA 93117, USA}

\author{P. Roushan}
\affiliation{Google Inc., Santa Barbara, CA 93117, USA}

\author{D. Sank}
\affiliation{Google Inc., Santa Barbara, CA 93117, USA}

\author{A. Vainsencher}
\affiliation{Google Inc., Santa Barbara, CA 93117, USA}

\author{J. Wenner}
\affiliation{Department of Physics, University of California, Santa Barbara, California 93106-9530, USA}

\author{T.C. White}
\affiliation{Department of Physics, University of California, Santa Barbara, California 93106-9530, USA}

\author{H. Neven}
\affiliation{Google Inc., Santa Barbara, CA 93117, USA}

\author{John M. Martinis}
\email{jmartinis@google.com}
\affiliation{Department of Physics, University of California, Santa Barbara, California 93106-9530, USA}
\affiliation{Google Inc., Santa Barbara, CA 93117, USA}

\author{A. Megrant}
\email{amegrant@google.com}
\affiliation{Google Inc., Santa Barbara, CA 93117, USA}
\date{\today}

\begin{abstract}

Complex integrated circuits require multiple wiring layers.  In complementary metal-oxide-semiconductor (CMOS) processing, these layers are robustly separated by amorphous dielectrics.  These dielectrics would dominate energy loss in superconducting integrated circuits.  Here we describe a procedure that capitalizes on the structural benefits of inter-layer dielectrics during fabrication and mitigates the added loss.  We use deposited inter-layer dielectric throughout fabrication, then etch it away post-fabrication.  This technique is compatible with foundry level processing and can be generalized to make many different forms of low-loss wiring.  We use this technique to create freestanding aluminum vacuum gap crossovers (airbridges).  We characterize the added capacitive loss of these airbridges by connecting ground planes over microwave frequency $\lambda/4$ coplanar waveguide resonators and measuring resonator loss.  We measure a low power resonator loss of $\sim 3.9 \times 10^{-8}$ per bridge, which is 100 times lower than dielectric supported bridges.  We further characterize these airbridges as crossovers, control line jumpers, and as part of a coupling network in gmon and fluxmon qubits.  We measure qubit characteristic lifetimes ($T_1$'s) in excess of 30 $\mu$s in gmon devices.

\end{abstract}
\maketitle

Two dimensional superconducting qubit architectures will require multi-layer wiring.\cite{brecht2016multilayer, harris2010experimental, lanting2014entanglement, foxen2017qubit, rosenberg20173d}  Multiple wiring layers are fundamental to standard integrated circuits to route signals past one another to individually address a two dimensional grid of elements.  Multi-layer wiring has been developed for superconducting circuits.\cite{tolpygo2015fabrication, nagasawa2014nb}  These wiring layers are seperated by deposited dielectrics, and while these processes offer robust large scale control, the amorphous dielectrics used (typically SiO$_2$) are quite lossy, with loss tangents $\tan \delta \approx 10^{-3}$.\cite{o2008microwave, quintana2014characterization}  We limit participation of similar dielectrics ($p_i < 10^{-3}$) to achieve state-of-the-art qubit quality factors ($Q_i>1\times10^6$).\cite{dunsworth2017characterization, dial2016bulk}  We have developed a method that benefits from the structural support of inter-layer dielectrics while mitigating the loss.  We use deposited dielectrics only as a scaffold to separate and stabilize different metal layers through aggressive fabrication steps, and then etch it away at the end of fabrication.  This process is compatible with standard CMOS processing, and provide an avenue toward scalable low-loss control wiring for a two dimensional grid of qubits.  While this method is quite general and can be applied to many forms of multi-layer wiring we demonstrate this technique by fabricating the simplest forms of multi-layer wiring: crossovers.  

\begin{figure}[h!]
\includegraphics{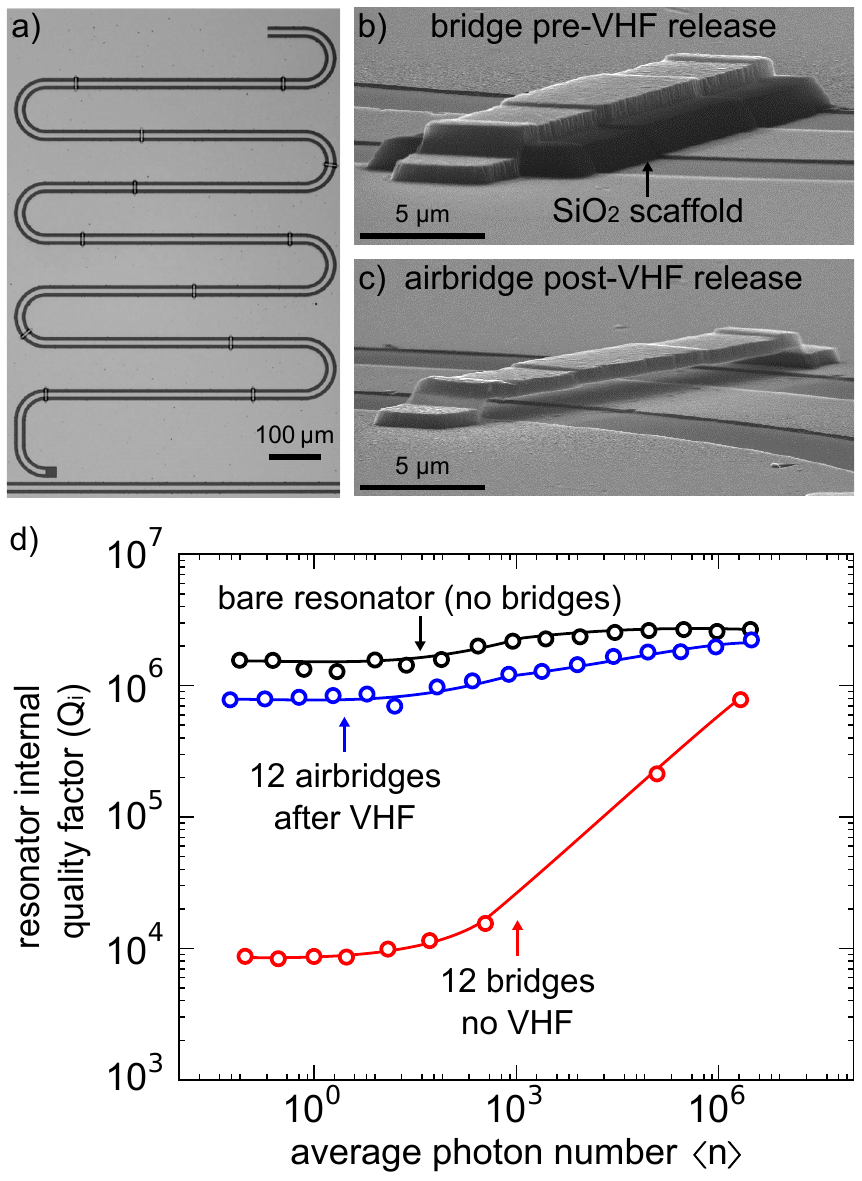} 
\caption{
(a) Optical micrograph of a CPW $\lambda/4$ resonator with 12 crossovers over the center trace capacitively coupled to a feed-line.
(b) Crossover spanning the resonator before removing SiO$_2$ scaffold.
(c) Freestanding airbridge crossover after VHF treatment.
(d) Representative resonator $Q_i$ vs average photon excitation.  Leaving the SiO$_2$ under the bridges greatly reduces the resonator's quality, as would most deposited dielectrics.  After removing the SiO$_2$ the resonator's quality recovers to about a factor of two lower than bare witness resonators.  The twelve evenly spaced airbridges only cover $\sim 0.7 \%$ of the $\lambda$/4 resonator geometry.  Resonators without bridges (bare resonators) show no substantial difference in quality with or without VHF treatment.
}
\label{figure:resonator} 
\end{figure}

	Free standing metallic crossovers, known as airbridges, are widely used in low-loss microwave circuits \cite{koster1989investigations, kwon2001low} as well as superconducting circuits.\cite{ chen2014fabrication, abuwasib2013fabrication, lankwarden2012development} These airbridges are typically fabricated using re-flowed photoresist as a scaffold, which is removed immediately after bridge fabrication and prior to further processing.  Released airbridges typically cannot withstand the sonication widely used to remove surface contaminants. Additionally, without dielectric support, the mechanical strength of freestanding airbridges relies on an arched shape.  Airbridges with spans much larger than their arched height tend to bend under the pressure of resist spins and bakes.  Thus, airbridges are made taller to span longer distances. Bridge height is limited by future processing, as standard high-resolution resists ($\sim 1$-$10 \ \mu$m thick) fail to protect taller airbridges from aggressive processing steps such as ion etching or lift-off.  We use our dielectric scaffolding technique to create a different kind of airbridge.  The dielectric scaffolding stabilizes these bridges through aggressive sonication and resist coating, thus decoupling the air bridges' span from it's height.  Mechanical tests indicate these airbridges span distances of at least 70 $\mu$m reliably.  The added capacitive loss per bridge is comparable to photoresist scaffolded airbridges and is $\sim 100\times$ less lossy than conventional dielectric crossovers (bridges with the scaffolding left in-tact, as in Fig.  \ref{figure:resonator}).
	
	We fabricate these airbridges after defining aluminum basewiring on high resistivity ($>$10 k$\Omega\cdot$cm) intrinsic (100) plane silicon substrates.  We optically pattern a tri-layer \cite{supplement} stack of resist as a lift-off mask and electron beam (e-beam) deposit 1 $\mu$m of SiO$_2$ to define our dielectric scaffold.  Due to the growth conditions the SiO$_2$ sidewalls form an approximately 45$^\circ$ with the substrate (see Fig. \ref{figure:resonator}(b)).  Next, we reapply the same lift-off process to define the bridge itself, except prior to deposition, we use an in-situ 400 V, 0.8 mA/cm$^2$ argon ion mill to remove the exposed native aluminum oxide on the basewire.  This mill allows DC electrical contact between base-wire aluminum and the 600 nm thick airbridge aluminum.  After all further processing we use a dry VHF etcher (PRIMAXX \textregistered VHF Etch Release Technology) to release the airbridges by removing the scaffolding SiO$_2$.  The chamber is pumped low vacuum, and the die is heated to 45 Celcius on a 3 inch silicon carrier wafer.  A mixture of HF vapor, nitrogen, and ethanol is then bled into the chamber at a total pressure of 125 Torr (parameters in Table \ref{table:vhf_parameters}).  The scaffold SiO$_2$ and native oxide of the exposed silicon substrate are removed after 2 cycles of 15 seconds without breaking vacuum, as shown in Fig \ref{figure:resonator}(c).  Vapor phase release significantly reduces the mechanical strength required to overcome sticition, a common failure in microelectromechanical systems (MEMs) devices.\cite{van2002physical, maboudian1997critical}  This process does not attack other materials used in qubit fabrication including aluminum, aluminum oxide, and silicon.

\begin{table}[h!]
\centering

\begin{tabular}{ c | c | c}
VHF Flow & N$_2$ Flow & Ethanol Flow \\
(SCCM) & (SCCM) & (SCCM) \\
\hline
190 & 1425 & 210
\end{tabular}

\caption{VHF etch parameters.}
\label{table:vhf_parameters}
\end{table}
	
	In our superconducting circuits these airbridges serve two main functions: `jumper airbridges' which hop lines over each other and `ground plane airbridges' which connect ground planes over lines.  Jumper airbridges hop circuit elements over each other for stronger couplings, smaller footprints, and design flexibility. These SiO$_2$ scaffolded airbridges can be made with contact pads as small as $1 \ \mu$m$^2$ and allow even micron width lines to hop over each other.  Ground plane airbridges are commonly used to electrically connect ground planes to suppress parasitic microwave frequency slot line modes which modify couplings and act as qubit loss channels in coplanar waveguide (CPW) geometries.\cite{ponchak2005excitation, houck2008controlling}  These airbridges also route return currents to reduce unwanted cross-talk between control lines.  	
	
	We measure the added capacitive loss from airbridges using $\lambda/4$ CPW resonators.  To measure resonator loss, we cool down chips in a heavily filtered\cite{barends2011minimizing} adiabatic demagnetization refrigerator with a base temperature of 50 mK.  We extract resonator internal quality factor (loss = $1/Q_i$) by measuring and fitting the microwave scattering parameters versus frequency near resonance.\cite{megrant2012planar} Each chip has ten resonators capacitively coupled ($Q_c$  between $5\times 10^5$ and $1\times 10^6$) to a common feedline.  These resonators have between zero and ninety-eight groundplane airbridges spanning their center trace. The airbridges are 3 $\mu$m wide and have a height above the center trace set by the original dielectric thickness of 1 $\mu$m.  In Fig. \ref{figure:resonator}(a) we show one such resonator resonator spanned by 12 ground plane airbridges equally spaced along the resonator after the coupling arm.  All resonators have a 10 $\mu$m center trace and a 5 $\mu$m gap to ground on either side, and resonance frequencies near 6 GHz.  
	
	We compare loss between three styles of resonators: resonators spanned by scaffolded bridges (Fig. \ref{figure:resonator}(b)), resonators spanned by airbridges (after VHF release, Fig. \ref{figure:resonator}(c)), and pureley CPW resonators with no crossovers of any kind (bare resonators).  In Fig. \ref{figure:resonator}(d) we display internal quality factor data for these three resonators.  For clarity, we show only a single representative trace from each. The single photon loss limit approximately captures the physics of energy loss in superconducting qubits at the same frequency.  The bare witness resonator has a low power internal quality factor of around $1.5 \times 10^6$ which is consistent with single layer fabrication resonators of the same geometry.  We saw little to no difference in bare resonator quality factors between chips with or without the VHF process.  When the SiO$_2$ is left intact, (as it would be in typical dielectric crossovers) the low power $Q_i$ drops to around $1 \times 10^4$.  This is consistent with an amorphous SiO$_2$ loss tangent of $\tan \delta \approx 10^{-3}$ and a participation of $10\%$ (roughly the added capacitance for twelve scaffolded bridges).  After the VHF treatment, the $Q_i$ of resonators with twelve airbridges recovers to a factor of 2 lower than the bare resonators.  We measure the scaling of this residual loss with number of airbridges between zero and ninty-eight. The internal quality factor decreases with increasing number of airbridges and lines of best fit indicate added loss at low power of 3.9$\times 10^{-8}$ per bridge.\cite{supplement}

\begin{figure}[h!]
\begin{centering}
\includegraphics{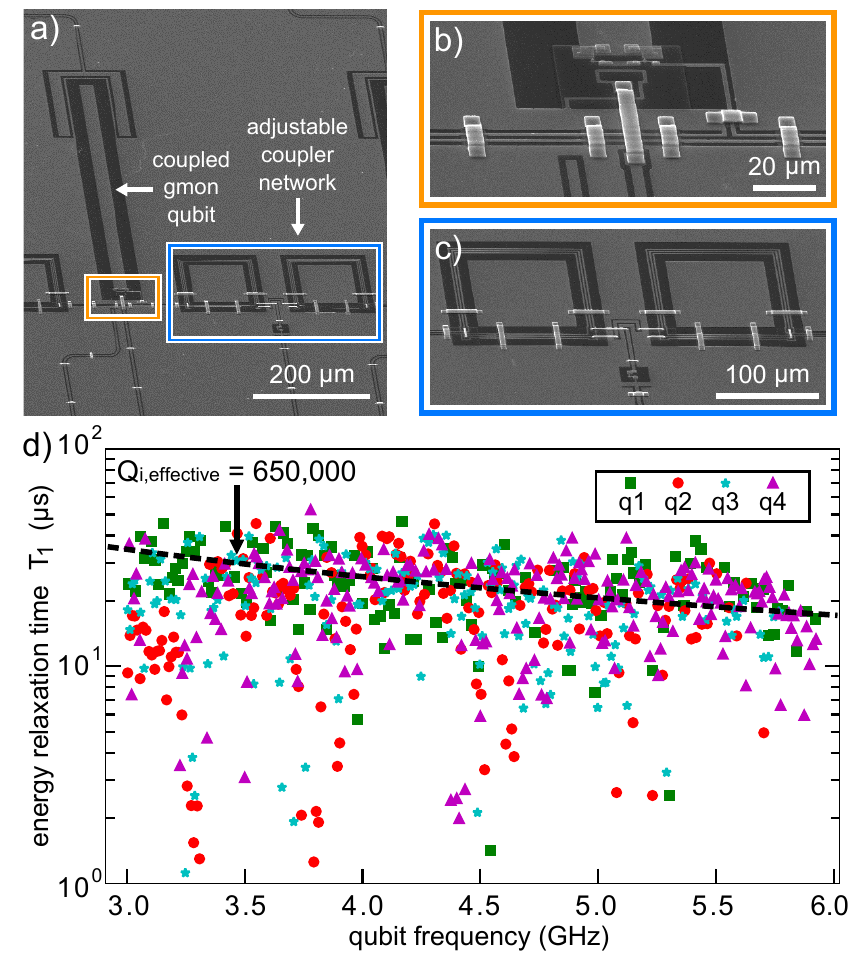}
\par\end{centering}
\caption{
(a) Scanning electron microscope image of a gmon qubit and it's neighboring adjustable coupler network.  
(b) This qubit design utilizes both jumper airbridges (hopping SQUID bias lines over ground plane and qubit inductor lines) as well as ground plane airbridges (hopping ground plane over coupler and qubit inductor lines).
(c) Jumper airbridges are also used in the coupler network to hop the coupler inductor over itself as well as qubit inductor lines creating a ``figure 8'' pattern.
(d) Qubit $T_1$ measurements from four different gmon qubits, with an average around 20 $\mu$s.}
\label{figure:gmon}
\end{figure}

	We use these low-loss airbridges as an integral part of gmon qubits. These qubits are transmon qubits \cite{koch2007charge} with inductive taps placed between DC SQUID and ground plane to allow adjustable coupling to nearest neighbors.\cite{chen2014qubit} It is critical that any added loss from the airbridges does not compromise the qubit coherence.  In Fig. \ref{figure:gmon}(a) we display one such gmon qubit with its neighboring coupler network.  We bias these qubits' DC SQUID loop with maximum DC current of 2 mA.  This current flows entirely through a jumper crossover in-line with the flux bias line (Fig. \ref{figure:gmon}(b)) and shows no evidence of on-chip heating.  In the qubit circuit, we use many ground plane airbridges as well as a set of jumper airbridges in-line with the coupler's geometric inductor (Fig. \ref{figure:gmon}(c)).  This jumper airbridge allows a gradiometric turn which further reduces crosstalk.  These jumper airbridges are only 1.5 $\mu$m wide, highlighting their small footprint.  It is also important to note that these airbridges are fabricated prior to Josephson junction deposition, and are robust after all of the further processing, with yield limited by errors in lithography.  In Fig. \ref{figure:gmon}(d) we show qubit energy relaxation time ($T_1$) spectra over 3 GHz of tunable qubit frequency for four different qubits.  The spectrum is well represented by a constant effective $Q_i \approx 6.5\times10^5$, with small sections where the $T_1$ drops dramatically.  These spectra are consistent with qubit loss dominated by dielectric surface loss from the SQUID area.\cite{dunsworth2017characterization} The airbridges themselves do not appear to greatly impact the qubit $T_1$ spectra.\cite{supplement}

\begin{figure}[h!]
\begin{centering}
\includegraphics{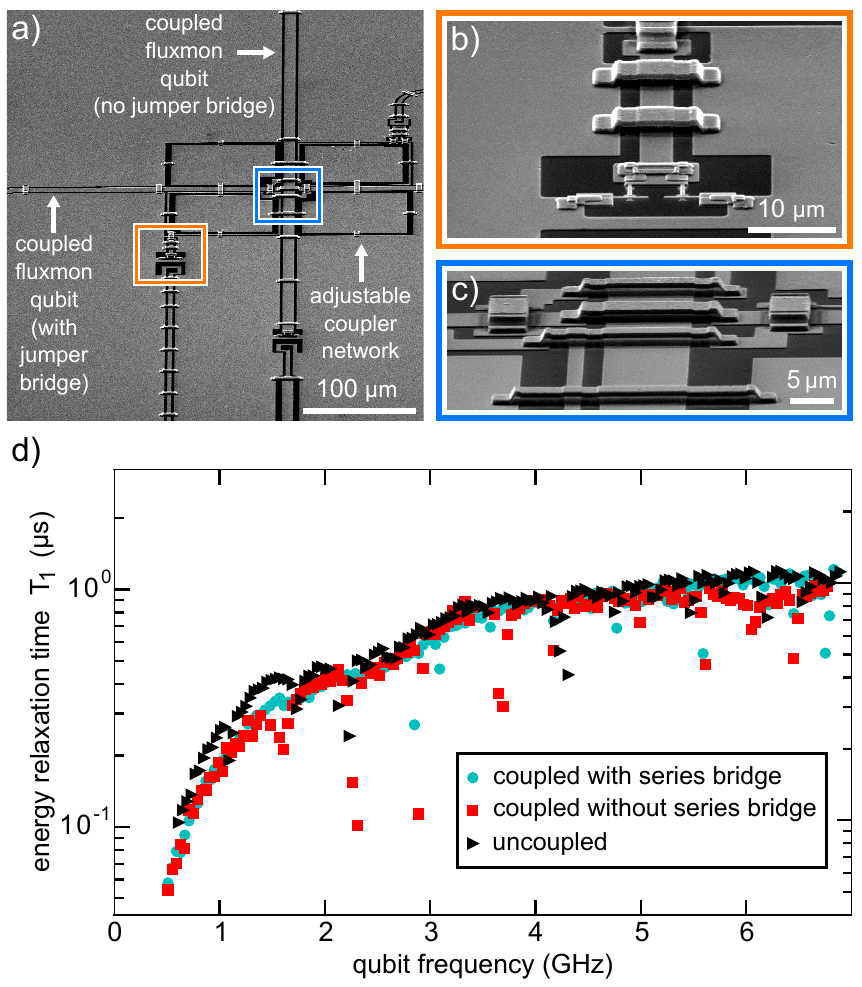} 
\par\end{centering}
\caption{
(a) Scanning electron microscope image of two coupled fluxmon qubits and their adjustable coupler.
(b) Image of ground plane and jumper airbridges near the coupler DC SQUID loop.
(c) Image of a network of airbridges with SiO$_2$ scaffolding removed after VHF processing.
(d) Qubit $T_1$ measurements from three different fluxmon qubits.  We see no systematic difference between coupled with series bridge, coupled with no series bridge, and uncoupled, indicating airbridges do not impact the coherence.}
\label{figure:fluxmon} 
\end{figure}

We also use these airbridges as integral parts of our fluxmon\cite{quintana2017observation} flux qubit circuits, for both isolated and coupled qubits.  The main inductance and capacitance of the fluxmon is distributed over a long CPW segment that is terminated with an electrical short to ground at one end and a DC SQUID shorted to ground at the other. We use both ground plane airbridges over the qubit's CPW and jumper airbridges in-line with the qubit's CPW and the couplers as well.   We tested three variations of fluxmon qubits on the same chip: uncoupled, coupled with jumper bridge, and coupled without jumper bridge. The uncoupled qubits only use ground plane airbridges. For the coupled qubits, the CPW center trace of one qubit jumps over the CPW center trace of the other qubit via an airbridge, as shown in Fig. \ref{figure:fluxmon}(a), while the other qubit does not have any in-line jumper airbridges.

The resulting qubit $T_1$ vs. frequency at symmetric bias (zero tilt bias) is shown in Fig. \ref{figure:fluxmon}(d) for the three qubit variations. The background dissipation is believed to come from $1/f$ flux noise at low frequencies\cite{quintana2017observation, yan2016flux} extrinsic to the airbridges, with some other inductive loss extrinsic to the airbridges dominating at high frequencies. We find no measurable difference in coherence between the two types of coupled qubits. This is consistent with a very high quality galvanic contact between the jumper bridge and the qubit's CPW. Furthermore, we see no measurable difference in coherence between the coupled and uncoupled qubits, despite the fact that the coupled qubits are in very close proximity to a coupler circuit (the thin traces and ground plane pads in Fig. \ref{figure:fluxmon}(c)) containing many crucial jumper and ground plane airbridges. This retained coherence is very important for scaling up fluxmon circuits with many jumper airbridges and couplers, in order to couple one qubit to many others at once for quantum annealing applications.

In summary, we have demonstrated a procedure that utilizes the structural benefits of inter-layer dielectrics commonly used in multi-layer wiring, while mitigating the capacitive loss.  We use this process to fabricate low-loss airbridges that are robust during fabrication against strong sonication, other aggressive etches, and have a low profile.  We measure the added loss per ground plane bridge over resonators to be $\sim 3.9 \times 10^{-8}$ at low power.  We demonstrated these airbridges' use in different superconducting qubit devices and measured little to no effect on the coherence of the qubits.  These qubit designs fundamentally require a second layer of wiring, and here we have demonstrated a proof-of-principle method for rigidly scaffolding this second layer of wiring.  Reapplication of the lift-off steps of SiO$_2$ and metal (prior to VHF release) would allow for further layers of wiring as well as further complexity.\cite{supplement}  By replacing the lift-off steps in the bridge fabrication with more standard blanket depositions and via etches, this technique is completely compatible with standard multi-layer CMOS processing.  

\section{Acknowledgments}
This work was supported by Google. C. Q. and Z.C. acknowledge support from the
National Science Foundation Graduate Research Fellowship under Grant No. DGE-
1144085. Devices were made at the UC Santa Barbara Nanofabrication Facility, a
part of the NSF funded National Nanotechnology Infrastructure Network. 

\bibliographystyle{apsrev}
\bibliography{Scaffold_Bridges}
\end{document}


\author{A. Dunsworth}
\affiliation{Department of Physics, University of California, Santa Barbara, California 93106-9530, USA}

\author{R. Barends}
\affiliation{Google Inc., Santa Barbara, CA 93117, USA}

\author{Yu Chen}
\affiliation{Google Inc., Santa Barbara, CA 93117, USA}

\author{Zijun Chen}
\affiliation{Department of Physics, University of California, Santa Barbara, California 93106-9530, USA}

\author{B. Chiaro}
\affiliation{Department of Physics, University of California, Santa Barbara, California 93106-9530, USA}

\author{A. Fowler}
\affiliation{Google Inc., Santa Barbara, CA 93117, USA}

\author{B. Foxen}
\affiliation{Department of Physics, University of California, Santa Barbara, California 93106-9530, USA}

\author{E. Jeffrey}
\affiliation{Google Inc., Santa Barbara, CA 93117, USA}

\author{J. Kelly}
\affiliation{Google Inc., Santa Barbara, CA 93117, USA}

\author{P. V. Klimov}
\affiliation{Google Inc., Santa Barbara, CA 93117, USA}

\author{E. Lucero}
\affiliation{Google Inc., Santa Barbara, CA 93117, USA}

\author{J.Y. Mutus}
\affiliation{Google Inc., Santa Barbara, CA 93117, USA}

\author{M. Neeley}
\affiliation{Google Inc., Santa Barbara, CA 93117, USA}

\author{C. Neill}
\affiliation{Department of Physics, University of California, Santa Barbara, California 93106-9530, USA}

\author{C. Quintana}
\affiliation{Google Inc., Santa Barbara, CA 93117, USA}

\author{P. Roushan}
\affiliation{Google Inc., Santa Barbara, CA 93117, USA}

\author{D. Sank}
\affiliation{Google Inc., Santa Barbara, CA 93117, USA}

\author{A. Vainsencher}
\affiliation{Google Inc., Santa Barbara, CA 93117, USA}

\author{J. Wenner}
\affiliation{Department of Physics, University of California, Santa Barbara, California 93106-9530, USA}

\author{T.C. White}
\affiliation{Department of Physics, University of California, Santa Barbara, California 93106-9530, USA}

\author{H. Neven}
\affiliation{Google Inc., Santa Barbara, CA 93117, USA}

\author{John M. Martinis}
\email{martinis@physics.ucsb.edu}
\affiliation{Department of Physics, University of California, Santa Barbara, California 93106-9530, USA}
\affiliation{Google Inc., Santa Barbara, CA 93117, USA}

\author{A. Megrant}
\email{amegrant@google.com}
\affiliation{Google Inc., Santa Barbara, CA 93117, USA}

\title{Supplementary Material for ``A Method for Building Low Loss Multi-Layer Wiring for Superconducting Microwave Devices"}

\date{\today}

\begin{abstract}

We provide supplementary data and calculations.

\end{abstract}

\maketitle

\begin{table*}[!ht]
\centering
\begin{tabular}{|c|c|c|c|c|c|}
\hline
\begin{tabular}[c]{@{}c@{}}Resist\\ Type\end{tabular} & \begin{tabular}[c]{@{}c@{}}Hot Plate Bake\\  Temperature (C)\end{tabular} & \begin{tabular}[c]{@{}c@{}}Bake Time \\ (minutes)\end{tabular} & \begin{tabular}[c]{@{}c@{}}Approximate \\ Thickness (nm)\end{tabular} & \begin{tabular}[c]{@{}c@{}}Vertical Oxygen \\ Barrel Ash Rate\footnote[1]{0.3 torr O$_2$, 100 watts RF bias power}\\ (nm/sec)\end{tabular} & \begin{tabular}[c]{@{}c@{}}Vertical Oxygen\\ Ash Rate in ICP\footnote[2]{0.015 torr O$_2$, 100 watts ion power, 0 watts RF bias power} \\ (nm/sec)\end{tabular} \\ \hline
\textbf{PMMA} & 160 & 10 & 240 & 2.10 & 0.92 \\ \hline
\textbf{PMGI} & ~ & ~ & ~ & ~ & ~ \\
SF5 (etch) & 160 & 5 & 200 & NA & NA \\ 
SF11 (liftoff) & 160 & 5 & 1300 & NA & NA \\ \hline
\textbf{SPR (955-0.9)}& 90 & 1.5 & 900 & 0.71 & 0.40 \\ \hline
\end{tabular}
\caption{Parameters for resists used in tri-layer stack.  All resists are spun on at 1500 rpm for 45 seconds.  SF5 is used for etch processes while the thicker SF11 is used for liftoff processes. We use a 0.4 second exposure at $\sim$420 mW/cm$^2$ at the wafer to expose the SPR, and do a post exposure bake on a 110 C hot plate for 90 seconds to improve resist contrast and development stability. Etch rates measured with blanket films of the corresponding resist types.}
\label{table:tri-layer}
\end{table*}

\section{resonator loss per bridge}

\begin{figure}[!ht]
\begin{centering}
\includegraphics{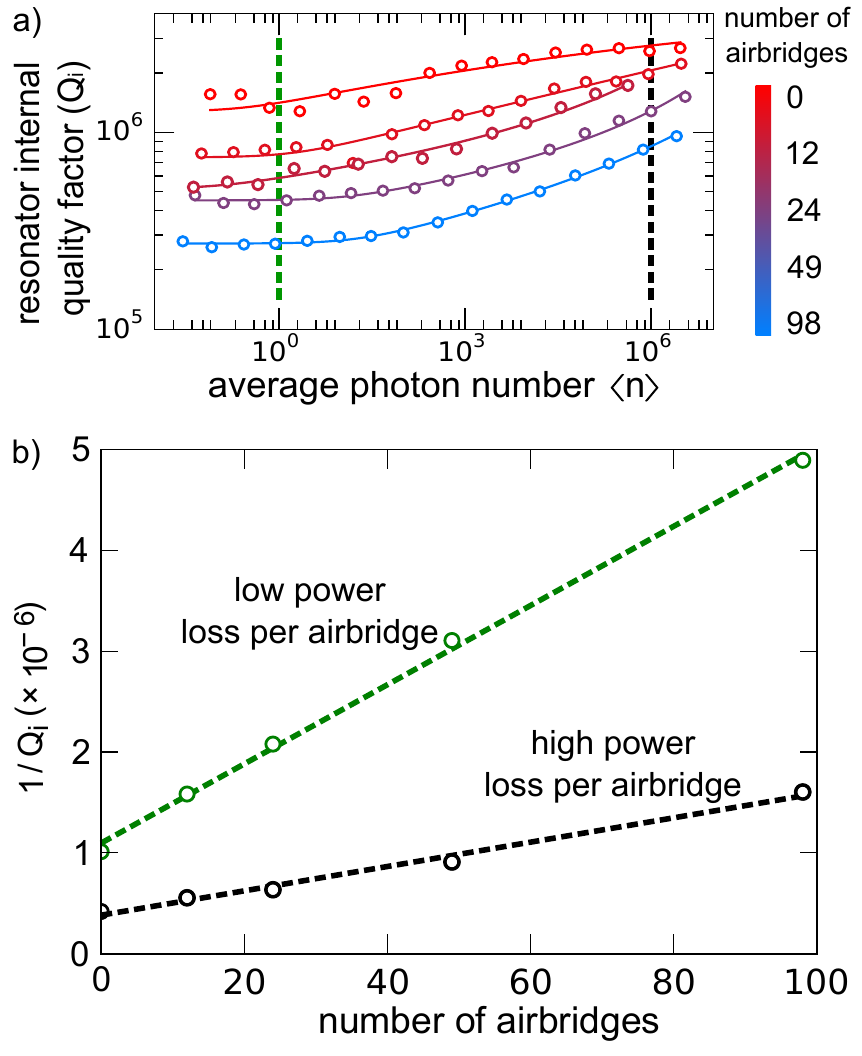}
\par\end{centering}
\caption{
Resonator loss (1/$Q_i$) cuts at low power (average photon population of  $\sim 10^0$) and high power (average photon population of $\sim 10^6$) plotted against number of bridges.  Lines of best fit give and $3.9 \times 10^{-8}$ ($1.2 \times 10^{-8}$) loss per bridge at low (high) power.}
\label{figure:resonator_loss_per_bridge}
\end{figure}

We design $\lambda/4$ coplanar waveguide (CPW) resonators with a variable number of ground plane airbridges to measure the added capacitive loss per bridge.  We use the the anhydrous hydrofluoric acid vapor (VHF) process detailed in the main paper to remove the SiO$_2$ scaffold prior to cooling down these resonators.  We measure the scaling of the resonator loss with between 0 and 98 bridges spanning the center trace. In Fig. \ref{figure:resonator_loss_per_bridge} (a) we display representative $Q_i$ vs average photon excitation for these resonators.  The resonator internal quality factor decreases with increasing number of bridges. In Fig. \ref{figure:resonator_loss_per_bridge} (b) we show cuts of loss ($1/Q_i$) vs number of bridges at low and high power. A line of best fit indicates an added loss at low power of 1.2$\times 10^{-7}$ per $f$F of added capacitance, or 3.9$\times 10^{-8}$ per bridge at low power. This is a factor of two higher loss per added capacitance of photoresist scaffolded airbridges ($5.08 \times 10^{-8}$ per $f$F).\cite{chen2014fabrication}  It is also important to note that if either of these bridges were coupled to a lumped capacitor, they would display a factor of two more loss.  Here we are protected from the full loss by the cosine voltage profile along the $\lambda /4$ resonator.

\begin{figure}[!ht]
\begin{centering}
\includegraphics{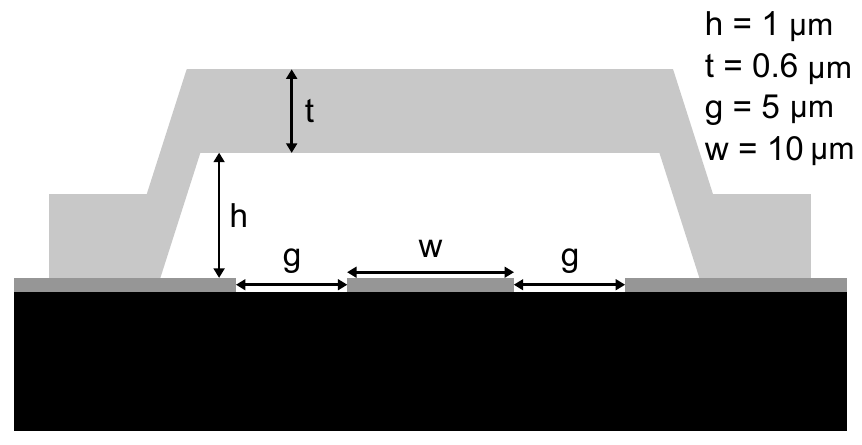}
\caption{CPW dimensions of aluminum resonators on a silicon substrate.  The width of the bridge in into the page is $l=3 \ \mu$m.}
\label{figure:bridge_dimension}
\end{centering}
\end{figure}

Here we calculate the expected added low-power loss per bridge:

$$1/Q_{i,bridge} = \tan \delta \times p_{loss}$$
$$\approx \tan \delta \left( \frac{2t_{loss}}{h} \right) \left( \frac{1}{\epsilon_{r,loss}} \right) \left(\frac{C_{bridge}}{C_{\lambda/4}}\right)$$
$$=1\times 10^{-9}  \ \frac{\mathrm{loss}}{\mathrm{nm}} \left( \frac{t_{loss}}{\epsilon_{r, loss}}\right)$$

Where the factor of 2 assumes that the lossy material is on both the top of the center conductor and bottom of the bridge equally.  We also assumed a loss tangent of $1\times10^{-3}$, consistent with previous works.\cite{wenner2011surface, o2008microwave}.  The capacitances are calculated as follows:

$$C_{\lambda/4}=\frac{1}{8 f_0 Z_0} \approx 470 \ \mathrm{fF}$$ 
$$C_{bridge}=\epsilon_0 \left( \frac{\mathrm{w}l}{\mathrm{h}} \right) \approx 0.266 \ \mathrm{fF}$$ 

Where we assume the geometries are all as displayed in figure \ref{figure:bridge_dimension}.  If we then assume the loss comes from the native oxide of aluminum, $t_{loss}=3 $ nm and $\epsilon_{loss}=10$, we get $3 \times 10^{-10}$ loss per bridge.  This greatly under predicts the loss.  If we assume it is left over SiO$_2$ ($\epsilon_{loss}=4$) it would require around 100 nm of lossy material to recover the above measured loss per bridge in this simple parallel plate model.  We do not see this thickness of residue in edge on SEMs similar to those in the main paper.

\section{Effect of Over-Etching SiO$_2$}

\begin{figure}[!ht]
\begin{centering}
\includegraphics{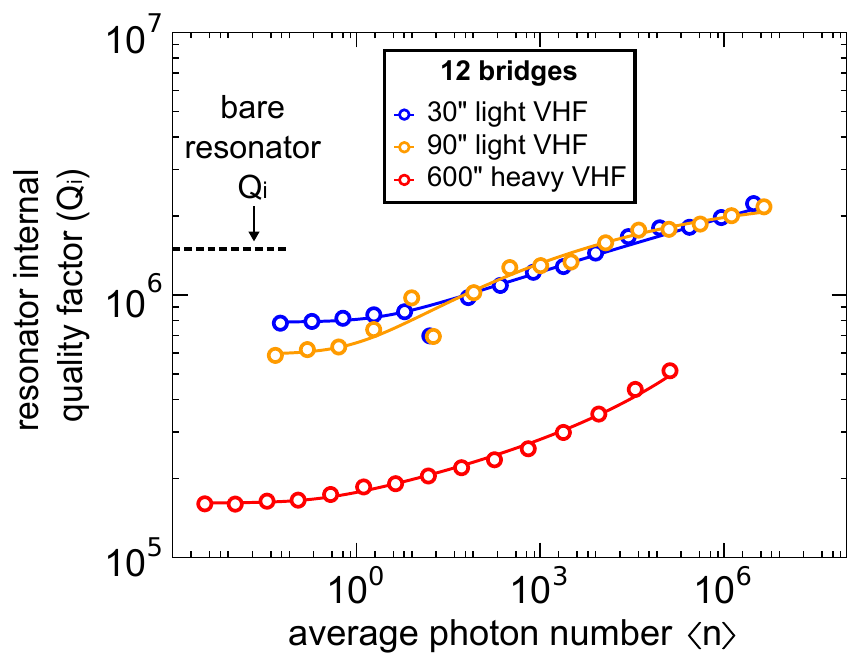}
\caption{Representative resonator $Q_i$ measurements for differing amounts of VHF treatment.  The bare resonator has no bridges, but did receive 30 seconds of the light VHF treatment detailed in the main paper.}
\label{figure:overetch}
\end{centering}
\end{figure}

The etch rate of the SiO$_2$ in VHF will depend on the amount of SiO$_2$ present.  This loading effect could lead to remnant SiO$_2$ and therefore increased loss.  Over-etching may also lead to excess loss, as VHF is known to leave residue from condensation under certain etch conditions.\cite{carter2000surface} We cooled down resonators etched for longer in VHF, as well as resonators with a much more substantial VHF etch (parameters in Tab. \ref{table:over_etch}), to test the effects of over-etching.  

\begin{table}[h!]
\centering
\begin{tabular}{ c | c | c}
VHF Flow & N$_2$ Flow & Ethanol Flow \\
(SCCM) & (SCCM) & (SCCM) \\
\hline
880 & 325 & 720
\end{tabular}
\caption{Heavy VHF etch parameters.}
\label{table:over_etch}
\end{table}

In Fig. \ref{figure:overetch} we plot $Q_i$ vs average photon population in resonators that underwent the above processes.  We note that there is a very small effect on the internal quality from over etching for up to 3 times the length required to remove the SiO$_2$ (this variation in $Q_i$ is expected for device-to-device variation).  However, when we use the stronger etch parameters for a much longer time, resonator's internal quality factor drops to around $2\times 10^5$.

\section{Full Fabrication}

\begin{figure}[h!]
\begin{centering}
\includegraphics{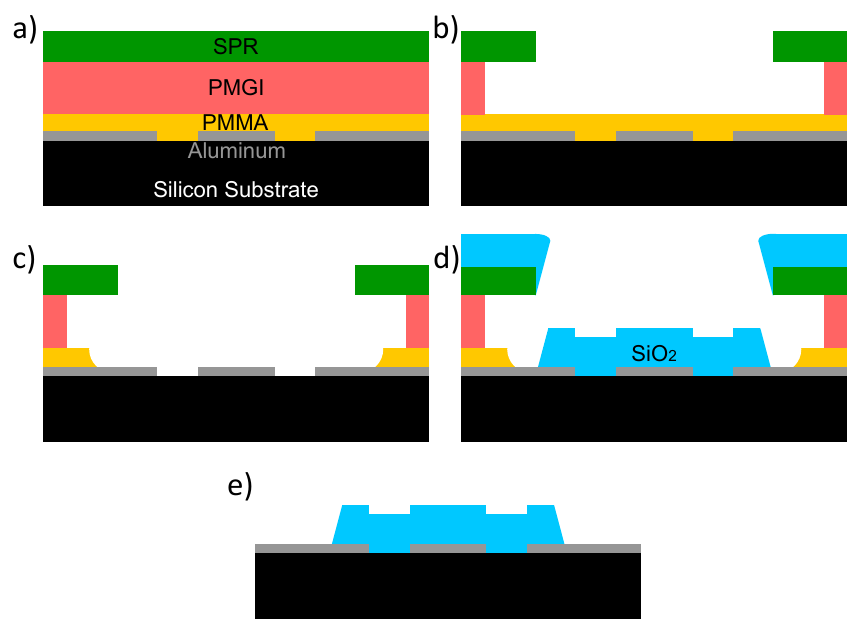} 
\par\end{centering}
\caption{Example tri-layer lift-off process step through.
(a) Cartoon profile of tri-layer stack of resists top-to-bottom SPR 955, PMGI, PMMA, then an etched 100 nm aluminum film on an intrinsic silicon substrate (not to scale).
(b) After photo lithography we develop to remove SPR where exposed and the PMGI develops isotropically at a rate of $\sim$2.4 $\mu$m/min.
(c) The PMMA is nearly directionally ashed (due to RF bias and low pressures) in an oxygen plasma.
(d) The SiO$_2$ is e-beam deposited in a high vacuum system.
(e) The resist is stripped clean with the help of the undercut layers breaking up the lift-off film.}
\label{figure:trilayer} 
\end{figure}

	The basewire deposition, lithography, and wet-etch (using Tetramethylammonium hydroxide (TMAH) based photo resist developer) are covered in detail in a previous publication. \cite{dunsworth2017characterization}  We use this tri-layer stack of resist, consisting of PMMA (Polymethylmethacrylate 4\% in Anisole), SF series PMGI, and i-line positive photoresist (SPR 955-0.9) to protect the aluminum from developer etching during dry etch and lift-off steps (Tab. \ref{table:tri-layer}).  The top resist layer is a standard photoresist for defining features $\sim 1 \ \mu$m in critical dimension.  The middle layer of resist allows for variable undercutting for reliable lift-off and and etching profiles.  The bottom layer of resist is used to protect the aluminum layer during photo resist development, and is known to etch readily in oxygen plasmas.\cite{quintana2014characterization, pop2012fabrication} We use a GCA Auto-Stepper 200 to expose optical patterns. The topmost resist layer develops where exposed.  The PMGI develops without being exposed, undercutting the SPR (Fig. \ref{figure:trilayer} a-b).  The PMMA is not etched by the TMAH based developer and thus protects the aluminum from being etched.  We then oxygen ash the PMMA to remove it where exposed and slightly undercut the SPR (Fig. \ref{figure:trilayer} c). 
	
	This tri-layer process is made compatible with both etching and lift-off processes by changing the PMGI thickness and PMMA ashing method.  For etch steps, the oxygen ash is done prior to the aluminum etch in-situ  in an inductive coupled plasma (ICP) tool (Panasonic E626I) with 15 mT of oxygen and 200 W plasma power with no RF bias onto the devices.  For lift-off steps we ash the PMMA in a barrel asher (Technics PEII) with 300 mT of oxygen with 300 W of RF power.  The middle PMGI layer also serves as a buffer between the solvents in the SPR and the PMMA.  Direct contact between SPR and PMMA leads to variable intermixing and unstable PMMA ash rates.  We use a thicker layer of PMGI SF11 ($\sim$ 1.1 $\mu$m) to fabricate the SiO$_2$ scaffolded airbridges bridges.  We do one round of lithography and ashing, then load into a high vacuum electron beam deposition tool (base pressure $\sim 1 \times 10^{-6}$ Torr) and deposit 1 $\mu$m of silicon oxide (Fig. \ref{figure:trilayer} d).  The resist is stripped using an N-Methyl-2-pyrrolidone (NMP) based resist stripper lifting off the excess silicon oxide (Fig. \ref{figure:trilayer} e), and a second round of spins and photo-lithography defines the top metal.  We load into another electron beam deposition tool ($P_{base}=2 \times 10^{-7}$ Torr), do an in-situ argon ion mill to remove the oxide of the exposed aluminum. We use a 400 V, 0.8 mA/cm$^2$ beam for 6 minutes with and continuous Argon flow of 3.6 sccm for this clean.  We then deposit 600 nm of aluminum to form the bridge.  We strip the resist as above to lift-off the excess metal.
	
	We have greatly stabilized our lithography and processing by using this tri-layer stack of resists. Stripping resist after dry etch steps is more stable as all the resist in direct contact with the substrate and metal is shielded from the high energy ions needed to etch the aluminum oxide and subsequently the underlying aluminum.  This allows solvents to get under hardened resist and reduces residues.  This tri-layer of resist also greatly stabilizes lift-off processing.  The undercut of the resist disconnects the lifted off film from the intended remnant material. This stack up also allows for an arbitrary number of lithography steps to be performed without worry of developer etching aluminum. This protection enables quick recovery from errors in lithography.

\section{Other Structures / Scaling to More Layers}

\begin{figure}[h!]
\begin{centering}
\includegraphics{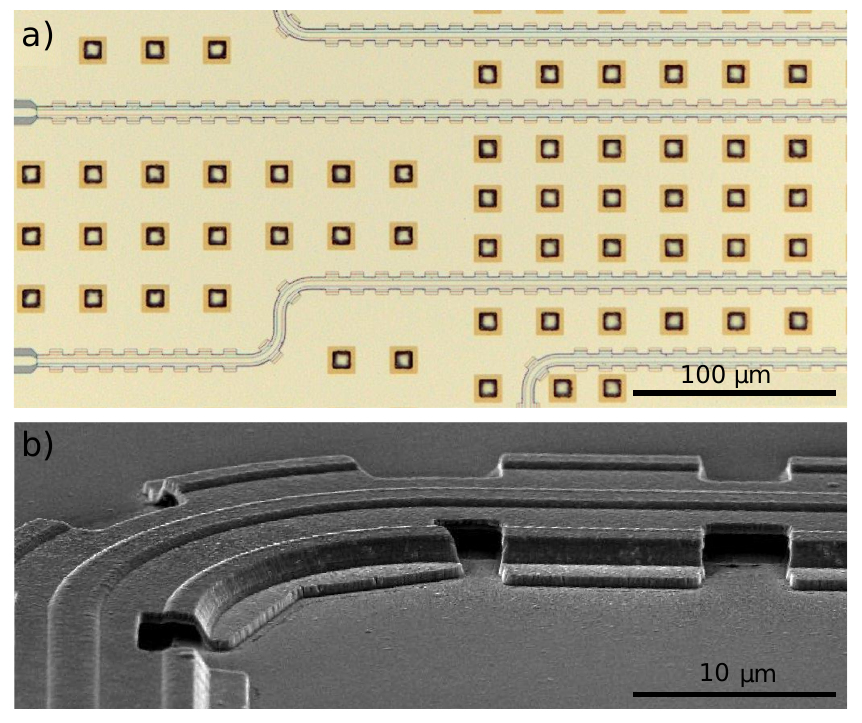}
\caption{
(a) Optical micrograph of CPW lines completely covered by a ``tunnel'' of metal.  
(b) The gaps between legs allow VHF to etch away the scaffolding SiO$_2$.}
\label{figure:tunnel}
\end{centering}
\end{figure}
 
 We use the same dielectric scaffolding technique (detailed in the main paper) to create a ``tunnel'' structure displayed in  Fig. \ref{figure:tunnel}.  This almost completely shields CPW lines from one side by covering it in a continuous ground plane.  This structure is useful for reducing cross-talk in sensitive devices.  It is perhaps easier to see that a third wiring layer could be fabricated on top of this tunnel.  We expect that a 10 $\mu$m center trace resonator that is entirely encased by a tunnel would have a low (high) power $Q_i$ of roughly $2 \times 10^4$ ($5 \times 10^4$) by extrapolating the above loss per airbridge (Fig. \ref{figure:resonator_loss_per_bridge} (b)).  However, these tunnel resonators may have a higher $Q_i$ than this extrapolation predicts, as the continuous covering would have no edges or corners that concentrate the electric field.  The $Q_i$ of these structures depends on the geometry.  Specifically, we chose the height of the tunnel (or bridge) to optimize for high Q as well as process stability for a single added layer.  These heights could be modified to more easily allow further layers to be added on top.
 
  It is important to note that the top metal layer is simply a re-application of the same lift-off steps as the scaffolding dielectric.  Therefore nothing fundamentally limits this process to only a second layer.  Furthermore, with planarization (a method commonly used in large layer stacks for CMOS processing\cite{nagasawa2014nb, tolpygo2015fabrication}), this process also generalizes to many multiple wiring layers.  Ventilation holes are required however (as in Fig \ref{figure:tunnel}) to allow the VHF to attack the underlying SiO$_2$.  While this constraint does add some complexity to the design layout, it is not prohibitive. If properly considered, this process  allows for even more complex wiring.  

\section{gmon $T_1$ frequency dependence}

\begin{figure}[h!]
\begin{centering}
\includegraphics{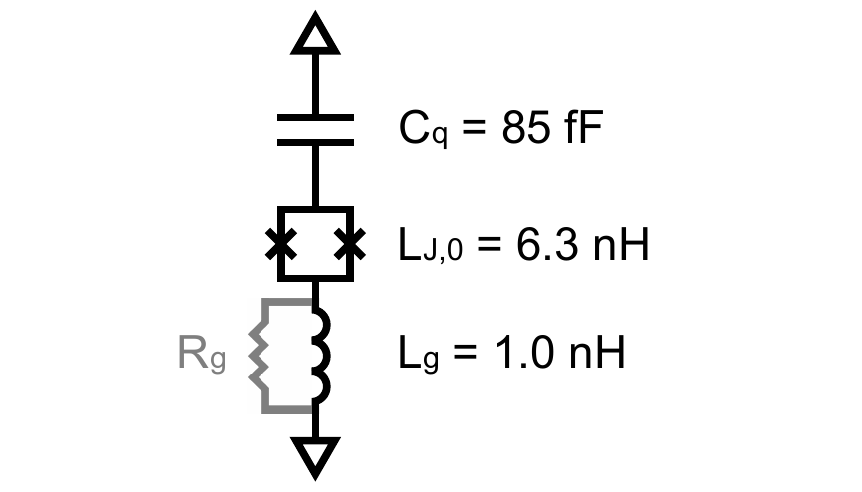}
\caption{A simplified circuit of a lone gmon.  The shaded resistor represents surface loss from the amorphous dielectrics near the thin geometric inductor lines.}
\label{figure:gmon_circuit}
\end{centering}
\end{figure}
	
	All of the bridges in the gmon circuit are most strongly coupled to the qubit's geometric inductor.  Therefore, additional loss from these bridges would mostly add to that of the stray capacitance of the geometric inductor.  The coherence of the gmon qubit is protected from capacitive loss in it's thin inductor lines by a voltage divider between it's SQUID inductance ($L_{J,0} \approx 6.3$ nH) and the linear geometric inductance ($L_g \approx 1.0$ nH).  We flux tune this SQUID inductance larger to decrease the qubit's frequency, thus the qubit energy relaxation time ($T_1$) would have a frequency dependence, as the ratio of SQUID to geometric inductance changes.  Here we calculate the expected frequency dependence of this loss.  It is often easier to think about loss in terms of an effective quality factor ($Q_i$).  For this channel:

\begin{equation}
Q_i = \frac{R_g}{Z_q}\left(\frac{V_q}{V_g} \right)^2
\end{equation}

Where $Z_q \approx (L_J/C_q)^{1/2}$ is the qubit impedance, $R_g$ is the loss from surface amorphous dielectrics near the geometric inductor lines, $V_q$ is the voltage drop across the qubit capacitor ($C_q$), and $V_R$ is the voltage drop across the geometric inductor tail ($L_g$).  We neglect the stray capacitance of the inductor tail as the qubit operates well below the resonance of the inductor circuit ($\sim 12$ GHz), and instead consider it only as a source of loss.  We can calculate $V_g$ in terms of $V_q$ using the voltage divider:

\begin{equation}
V_g = V_q\left(\frac{L_g}{L_J+L_g} \right) \approx V_q\left(\frac{L_g}{L_J} \right)
\end{equation}

we can also define $\omega_q=1/(L_J C_q)^{1/2}$ and thus:

\begin{equation}
Q_i \approx \frac{R_g}{C_q L_g^2} \left(\frac{1}{\omega_q}\right)^3
\end{equation}

and to convert to an energy relaxation limit $T_1=Q_i/\omega_q$ of the qubit:

\begin{equation}
T_1 = \frac{R_g}{C_q L_g^2} \left(\frac{1}{\omega_q}\right)^4
\end{equation}

	We do not witness this strong frequency dependence in the qubit's energy relaxation spectrum, indicating the qubit's coherence is not limited by this loss channel.  

	Another main loss channel for these qubits is due to surface dielectrics in the qubit capacitor.  We fabricate witness resonators (etched at the same time as the qubit capacitor, but cooled down separately) to investigate this limit on qubit coherence.  Witness resonators with a similar geometry have a much larger $Q_i \approx 3\times10^6$ indicating that the qubits' $T_1$ is not limited by the capacitor itself.  Most likely the gmon's $T_1$ is limited by interfacial amorphous dielectrics near the Josephson junction electrodes.\cite{dunsworth2017characterization}

\bibliographystyle{apsrev}
\bibliography{supplement}